\documentclass[preprint,5p,times,twocolumn,sort&compress]{elsarticle}

\usepackage{amssymb}

\journal{Nuclear Instruments and Methods in Physics Research A}

\usepackage{relsize}
\def\cesrta{{C{\smaller[2]ESR}TA}}

\begin{document}

\begin{frontmatter}



\title{Shielded button electrodes for time-resolved measurements of electron cloud buildup}



\author[Cornell]{J.A.~Crittenden\corref{cor1}}
\ead{crittenden@cornell.edu}
\author[Cornell]{M.G.~Billing}
\author[Cornell]{Y.~Li}
\author[Cornell]{M.A.~Palmer}
\author[Cornell]{J.P.~Sikora}
\address[Cornell]{CLASSE\fnref{fn1}, Cornell University, Ithaca, NY 14853, United States}
\fntext[fn1]{Work supported by the US National Science Foundation 
(\mbox{PHY-0734867}, \mbox{PHY-1002467}, and \mbox{PHY-1068662}), US Department of Energy
(\mbox{DE-FC02-08ER41538}), and the Japan/US Cooperation Program}
\cortext[cor1]{Corresponding author. Tel.: +1 6072554882}

\begin{abstract}
We report on the design, deployment and signal analysis
for shielded button electrodes sensitive to
electron cloud buildup at the Cornell Electron Storage Ring. These simple detectors, derived
from a beam-position monitor electrode design,
have provided detailed information on the physical processes underlying the local 
production and the lifetime of electron densities in the storage ring.
Digitizing oscilloscopes are used to record electron fluxes incident on the vacuum chamber
wall in 1024 time steps of 100~ps or more. The fine time steps provide a detailed
characterization of the cloud, allowing the independent estimation of processes contributing
on differing time scales and providing sensitivity to the characteristic kinetic
energies of the electrons making up the cloud. By varying the spacing and population of 
electron and positron beam bunches, we map the time development of the various
cloud production and re-absorption processes. The excellent reproducibility of the measurements
also permits the measurement of long-term conditioning of vacuum chamber surfaces. 
\end{abstract}


\end{frontmatter}


\section{Introduction}
The buildup of electron clouds (ECs) can cause instabilities and emittance growth in 
storage rings with positively charged beams. 
Low-energy electrons can be generated by ionization of residual gas, by beam particle loss and by 
synchrotron-radiation-induced photo-effect on the vacuum chamber walls.
These electrons can generate secondary electrons, particularly when accelerated to high
energy by the stored beam~\cite{ECLOUD12:Tue1815}. 
We report on studies performed in the
context of the Cornell Electron Storage Ring Test Accelerator ({\cesrta}) program~\cite{CLNS:12:2084}, 
an accelerator R\&D program for future low-emittance electron and positron storage rings.
The production of photoelectrons by synchrotron radiation is by far the dominant cause of electron 
cloud development at such high-energy storage rings~\cite{PRL75:1526}.
Many techniques  for measuring the EC density have been developed at {\cesrta}. 
One class of detectors samples the flux of cloud electrons 
on the wall of the beam-pipe. This paper describes the
use of a shielded button electrode (SBE)
as such an electron flux detector with sub-nanosecond time-resolving
capability. 
The SBE is sometimes 
referred to as a shielded-pickup~\cite{ECLOUD10:PST09} or a shielded button pickup~\cite{PRSTAB11:094401}.
We outline several experimental techniques based on the performance
of this type of detector to quantify cloud growth and decay mechanisms. 

\section{The Shielded Button Electrode Detector}
\begin{figure}[htb]
   \centering
   \includegraphics*[width=0.98\columnwidth]{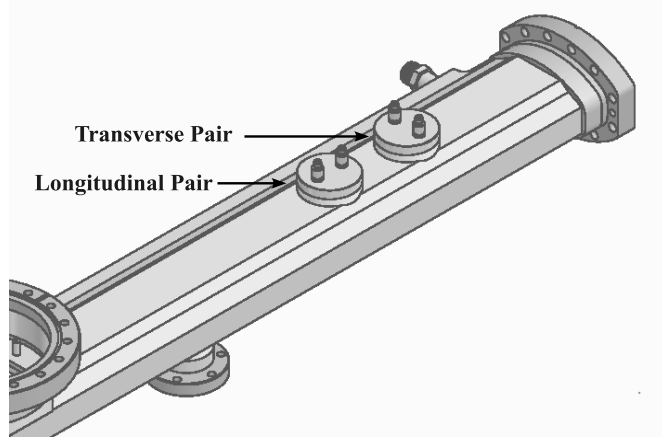}
   \caption{Custom vacuum chamber with shielded button electrodes.
            The SBEs, derived from beam-position monitor designs,
            are arranged in pairs: 
            one pair along the beam axis, the other pair transverse.}
   \label{SBE_beampipe}
\end{figure}

Two 1.1-m-long sections located
symmetrically in the east and west arc regions of the CESR ring were equipped with custom vacuum 
chambers
as shown in Fig.~\ref{SBE_beampipe}. A retarding-field analyzer port is shown on the left end,
and two SBE modules are shown near the right end of the
chamber, each with two detectors. 
The SBEs incorporate beam-position monitor (BPM) 
electrode designs, but placed outside the beam-pipe
behind a pattern of holes shielding them from the directly induced signal from the passing
beam bunches.
Two SBE electrodes are placed longitudinally, providing redundancy
and two others 
are arranged transversely, providing laterally segmented sensitivity to the 
cloud electrons. The centers of the latter two electrodes are ${\pm 14}$~mm from the
horizontal center of the chamber.

\begin{figure}[htb]
   \centering
   \includegraphics*[width=0.93\columnwidth]{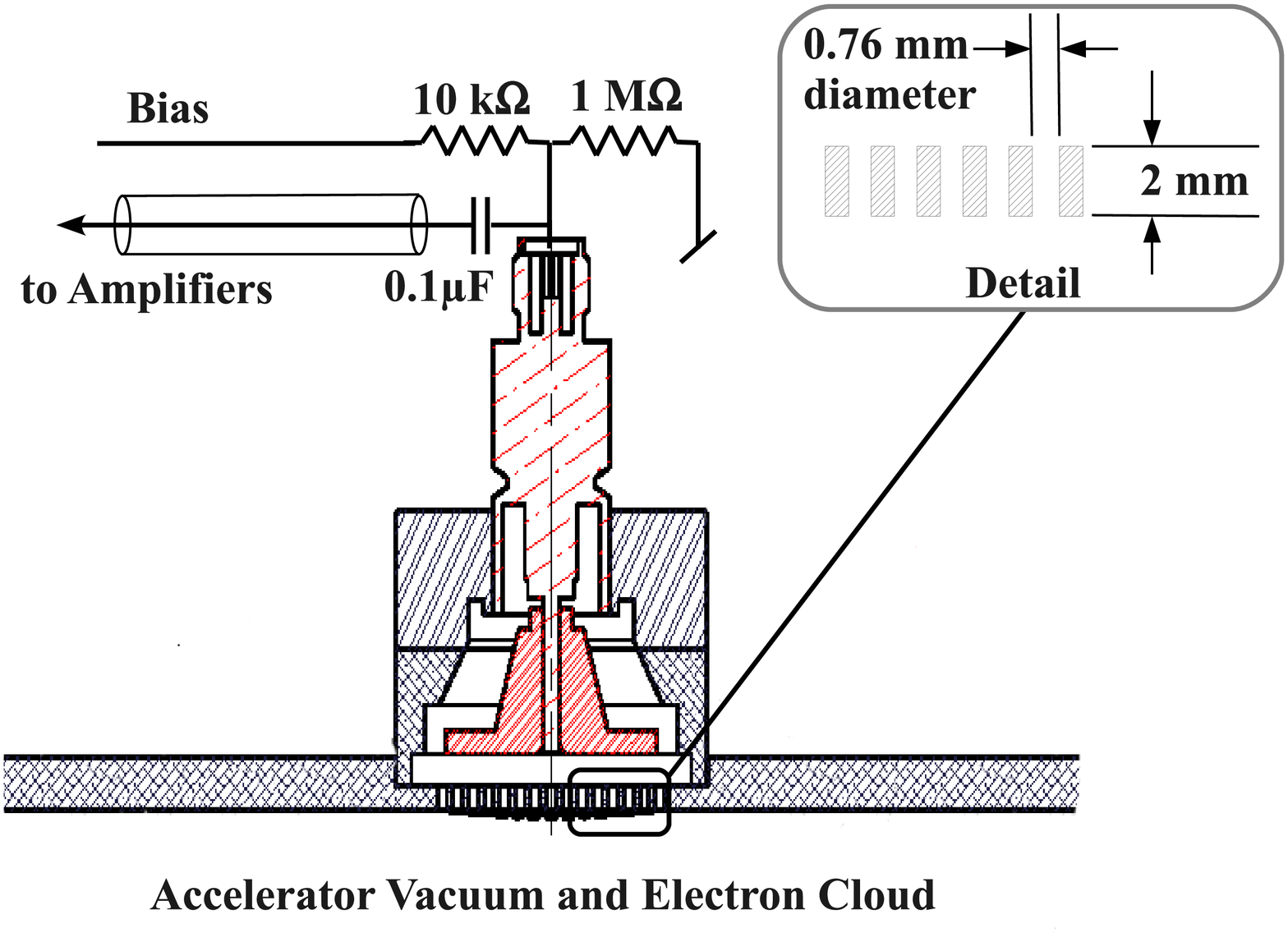}
   \caption{SBE detector design, biasing and readout. The 3:1 ratio of depth to diameter
            of the holes in the top of the beam-pipe effectively shields the 
            collector electrode from the direct beam signal. A 50-V positive
            bias serves to prevent secondary electrons produced on the electrode from escaping.}
   \label{SBE_Button}
\end{figure}

Figure~\ref{SBE_Button} shows schematically a cross-section of the SBE, 
the pattern of holes in the vacuum chamber allowing signal electrons to reach the button electrode, 
and the readout signal path. 
The distance from the beam-pipe surface to the electrode is 3~mm.
A DC bias relative to the grounded vacuum chamber is applied to the electrode through a 
10~k{$\Omega$} resistor. 
The signal is AC coupled to the 50~{$\Omega$} coaxial cable
through a 0.1~{$\mu$}F blocking capacitor which
provides high pass filtering.
A 1~M{$\Omega$} bleeder resistor provides a local ground path to prevent 
the electrode from charging up when the bias circuit is disconnected.
The front-end readout electronics comprise two Mini-Circuits ZFL-500 broadband amplifiers
with $50~{\Omega}$ input impedance for a total gain of 40~dB. Their bandwidth of 0.05-500~MHz
is approximately matched to the digitizing oscilloscope used to record their output signals.
Oscilloscope traces are recorded with 0.1~ns step size to 8-bit accuracy 
with auto-scaling, averaging over 8000 triggers. 
The fastest risetime recorded for EC signals has been less than 1~ns 
(see Sec.~\ref{sec:ecmeasurement}).
In contrast to the measurements provided by commonly used 
retarding-field analyzers~\cite{NIMA453:507to513,ARXIV:1402.1904}, which integrate the 
incident charge flux to provide a steady-state
signal current, our readout method provides time-resolved information on the cloud buildup, averaged
over 8000 beam revolutions in order to reduce sensitivity to asynchronous high-frequency noise. 
The trigger rate is limited by the oscilloscope averaging algorithm to about 1~kHz.
Since the beam revolution time is 2.5~{$\mu$}s, the cloud is sampled about once every 400 turns.

\begin{figure}[htb]
   \centering
   \includegraphics*[width=0.93\columnwidth]{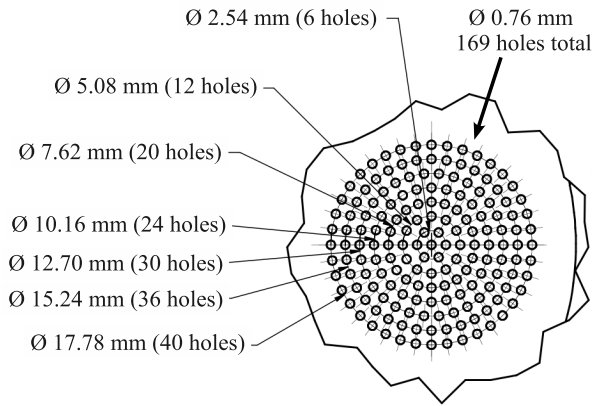}
   \caption{Hole pattern in the top of the vacuum chamber permitting signal electrons to reach the SBE. 
            The 169 holes are centered on seven concentric circles of diameters ranging from 2.54~mm to 17.78~mm.}
   \label{Holes_detail01}
\end{figure}

The hole pattern, shown in Fig.~\ref{Holes_detail01}, consists of 169 holes of 0.76~mm diameter arranged in 
concentric circles up to a maximum diameter of 18~mm. 
The hole axes are vertical. The approximate 3:1 depth-to-diameter factor is chosen to shield effectively
the detectors from the signal induced directly by the beam~\cite{PEP:253}.
The transparency for vertical electron trajectories is ~27\%. Together with the $1 \times 10^{-3}$~m$^2$ area
of the hole pattern, the 50~{$\Omega$} impedance and the 40~dB gain, this transparency results in a signal
of 1.35~V for a perpendicular current density of 1~A~m$^{-2}$.

A $50$~V positive bias on the button electrode serves to eliminate contributions to the signal 
from escaping secondary
electrons. Very few of these secondaries have kinetic energy sufficient
to escape a 50~V bias. This choice of bias also provides sensitivity to
cloud electrons which enter the holes in the vacuum chamber with low kinetic energy.

\section{\label{sec:ecmeasurement}
         Measurement of Electron Cloud Buildup Dynamics
        }
\begin{figure}[htb]
   \centering
   \includegraphics*[width=0.93\columnwidth]{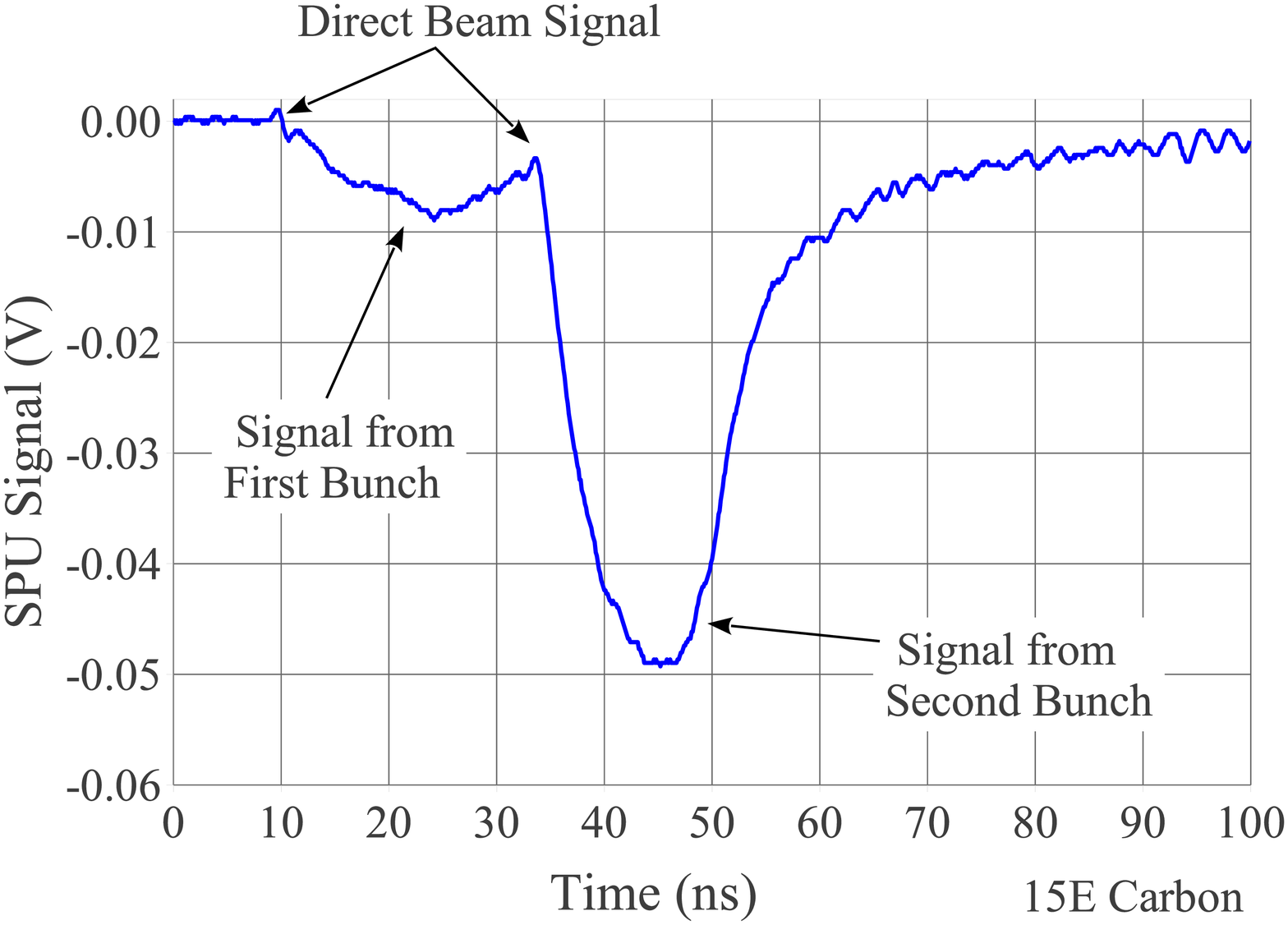}
   \caption{The SBE signal produced by two beam bunches 
            spaced by 24~ns, each comprising \mbox{$4.8{\times}10^{10}$} positrons.}
   \label{TwoBunchPositrons24ns}
\end{figure}

Figure~\ref{TwoBunchPositrons24ns} 
shows an example of a digitized SBE signal produced by two 5.3~GeV beam bunches each consisting of
\mbox{$4.8{\times}10^{10}$} positrons spaced 24~ns apart.  
The rms bunch length is 18~mm. Synchrotron
radiation of critical energy 3.8~keV from the upstream dipole magnet is absorbed on the vacuum 
chamber wall (amorphous-carbon-coated aluminum) nearly simultaneously with the arrival of the positrons. 
The arrival time of the 60-ps-long bunch is indicated by the small directly induced signal 
which penetrated the shielding holes, shown at a time of 10~ns in 
Fig.~\ref{TwoBunchPositrons24ns}. 
This small direct beam signal serves as a useful fiducial for determining the time interval between
bunch passage and cloud electron arrival times at the button electrode.

The time characteristics of such signals carry much detailed information on EC development. 
The leading bunch seeds the cloud and produces photoelectrons which can eventually pass into the SBE
detector. The signal from this first bunch is produced by the photoelectrons produced on the bottom of the 
vacuum chamber, since they are the first to arrive at the top of the chamber, accelerated 
by the positron bunch toward the detector above. The arrival times of the signal electrons
are determined by the combination of production energy, beam acceleration, and
the distance between the top and bottom of the vacuum chamber.
The second signal peak induced by the trailing (``witness'') bunch 
is larger, since it carries a contribution 
from the cloud present below the horizontal plane containing the 
beam when the bunch arrives.
Since these cloud 
electrons have been produced by wall interactions during the preceding 24~ns, the size and 
shape of 
this second signal peak depend directly on the secondary yield characteristics of the vacuum 
chamber surface. 

\begin{figure}[htb]
   \centering
   \includegraphics*[width=0.93\columnwidth]{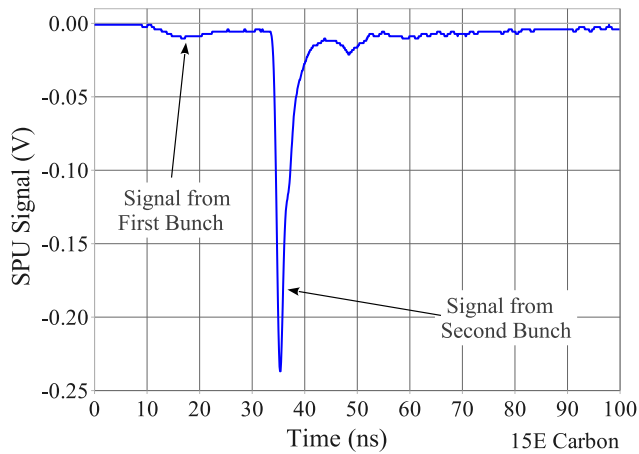}
   \caption{A pair of bunches consisting of  
            \mbox{$4.8{\times}10^{10}$} electrons spaced by 24~ns show a 
            dramatic difference in the first and second bunch signals similar to that observed for the
            positron bunches. 
            The  second bunch signal has a much faster rising edge than the 
            corresponding signal for a positron beam shown in Fig.~\ref{TwoBunchPositrons24ns}. }
   \label{TwoBunchElectrons24ns}
\end{figure}

Figure~\ref{TwoBunchElectrons24ns} shows the signals obtained from two electron bunches of 
similar length and population as the positron bunches considered above.
The primary source of synchrotron radiation is
of higher critical energy, 5.6~keV, since the source point is in a dipole magnet of 3~kG field,
rather than 2~kG. In addition, the incident photon rate is about a factor of three higher, since
the distance to the upstream dipole is 1~m rather than 3~m. 
The more dramatic difference between the signals from the first and second bunches
results from the fact that 
the witness-bunch signal arises from cloud electrons located above the horizontal plane containing
the beam at the bunch
arrival time, giving a much steeper risetime and a peak signal about five times higher. This
opposite beam kick also results in a signal of much shorter duration. The amplitude and time
dependence of the leading bunch signal are sensitive to the production kinetic energy distribution
of the photoelectrons, since they must overcome the beam kick in order to reach
the detector. Time-sliced numerical simulations have shown that such electrons
must be produced with hundreds of electron-volts of kinetic energy~\cite{ECLOUD10:PST09,PAC11:WEP142}. 
These photoelectrons, like the 
photoelectrons producing the lead signal with a positron beam, must be produced by synchrotron 
radiation which has undergone sufficient reflection to be absorbed on the bottom of the beam pipe.

\section{Measurement of Cloud Lifetime}
Such time-resolving measurements of the cloud evolution provide sensitivity to its
kinematic phase space distribution. The beam kicks, which can be controlled
by varying the bunch population,
accelerate cloud electrons to energies at and beyond the peak energy of the secondary emission
curve~\cite{PRSTAB5:124404}. Subsequent collisions with the vacuum chamber wall reduce the 
cloud kinetic energy. Eventually the secondary emission process is dominated by elastic reflection of the
remaining low-energy electrons. The cloud lifetime is then determined by the material-specific 
elastic yield value of the surface.

\begin{figure}[htb]
   \centering
   \includegraphics*[width=0.93\columnwidth]{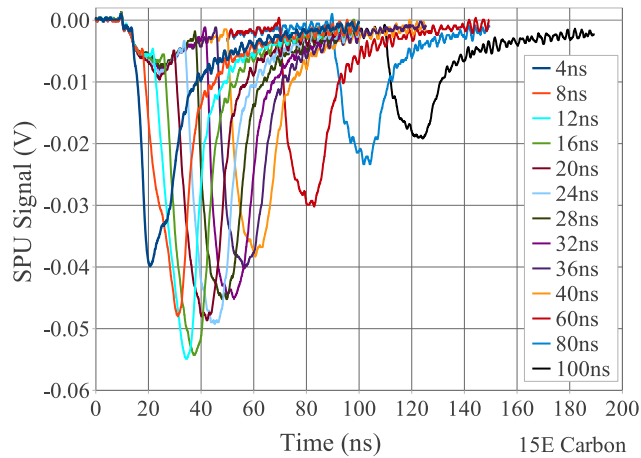}
   \caption{Overlay of thirteen 
            two-bunch signals with delays varying from 4 to 100~ns, including the 
            case of 24-ns delay shown in Fig.~\ref{TwoBunchPositrons24ns}. The time dependence of 
            EC buildup and decay are manifest. They result from the dependence 
            of the various secondary emission processes on the energies of cloud electrons
            colliding with the vacuum chamber surface.}
   \label{TwoBunchPositronsOverlay}
\end{figure}

Figure~\ref{TwoBunchPositronsOverlay} illustrates a method of determining cloud lifetime, 
and therefore the elastic yield value, for an amorphous-carbon-coating.
Overlaying the two-bunch
signals obtained by varying the delay in the arrival of the trailing bunch in 4-ns steps
clearly shows both the buildup and decay of the cloud density. 
The various secondary emission processes
contributing to buildup and decay~\cite{PRSTAB5:124404}
determine the delay which results in the maximum witness-bunch signal~\cite{ECLOUD12:Fri1240}.
For the  
\mbox{$4.8{\times}10^{10}$} bunch population shown here, the elastic yield property of the
surface dominates the signal decay rate at delays greater than about 60~ns. For smaller values
of the delay, the delay dependence of the witness-bunch amplitudes is governed by the
relationship between bunch spacing, cloud kinematics and the size of the vacuum chamber.  
Numerical simulations have shown the elastic yield value for such a carbon coating to be
less than 20\%, similar to that found for a titanium-nitride coating~\cite{ECLOUD12:Fri1240}.
In comparison, a similar study for an uncoated aluminum chamber found optimal agreement
with the measured witness-bunch signals for an elastic yield value of 40\%.

\begin{figure}[htb]
   \centering
   \includegraphics*[width=0.93\columnwidth]{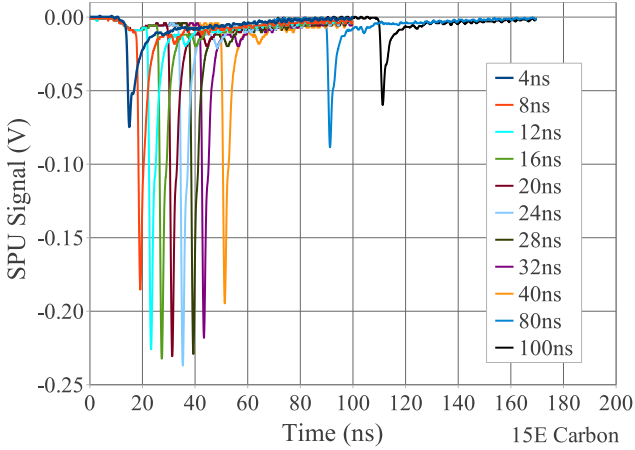}
   \caption{Overlay of eleven two-bunch signals with delays varying from 4 to 80~ns, including the 
            case of 24-ns delay shown in Fig.~\ref{TwoBunchElectrons24ns}. 
           }
   \label{TwoBunchElectronsOverlay}
\end{figure}

A similar witness-bunch study for an electron beam is shown in Fig.~\ref{TwoBunchElectronsOverlay}.
While the signals from each witness bunch differ from those obtained with a positron beam
as discussed in Sec.~\ref{sec:ecmeasurement}, the dependence on their delay times shows that detailed
information on cloud buildup and decay, with the attendant information on vacuum chamber
surface properties, can be obtained with an electron beam as well.


\section{Determination of Beam Conditioning Effects}
\begin{figure}[htb]
   \centering
   \includegraphics*[width=0.9\columnwidth]{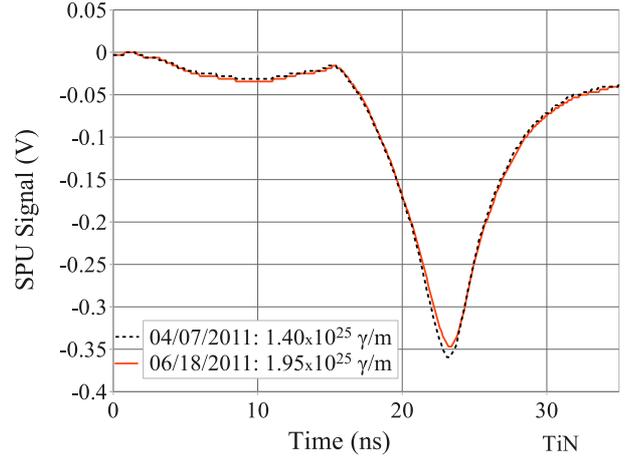}
   \vskip -1mm
   \caption{Comparison of SBE signals in April and June of 2011
            obtained from a pair of 5.3~GeV positron bunches
            of population \mbox{$8.2{\times}10^{10}$}
            separated by 14~ns. The change in the EC production
            properties of this TiN coating was negligible as the synchrotron radiation
            dose increased from \mbox{$1.4{\times}10^{25} \gamma$/m} 
            to \mbox{$1.95{\times}10^{25} \gamma$/m}. }
   \label{TwoBunchTiN}
\end{figure}

The assessment of electron-cloud mitigation techniques necessarily includes their variation 
with beam dose. The secondary emission yields of copper and aluminum surfaces are known 
to decrease dramatically with beam dose, while such an effect is known to be smaller
for TiN coatings~\cite{PAC11:TUP230}. The time-resolved measurements of the SBE
in the custom vacuum chambers of {\cesrta} provide accurate determinations of beam
conditioning effects owing to their reproducibility~\cite{IPAC11:WEPC135}. 
Figure~\ref{TwoBunchTiN} shows 
a comparison of two-bunch signals obtained in a TiN-coated aluminum chamber in April and 
June of 2011. 
During the intervening time period, CESR had operated as a high-current
light source, so the beam dose was high. Using the calculation of synchrotron radiation
power at this position in the ring, we convert from amp-hours to linear photon density to
obtain an increase in dose from \mbox{$1.4{\times}10^{25} \gamma$/m} 
to \mbox{$1.95{\times}10^{25} \gamma$/m} over this intervening period. 
The TiN-coating shows no change in its secondary yield over this time and the measured
two-bunch signals are reproducible at the level of a percent.

\begin{figure}[bht]
   \centering
   \includegraphics*[width=0.9\columnwidth]{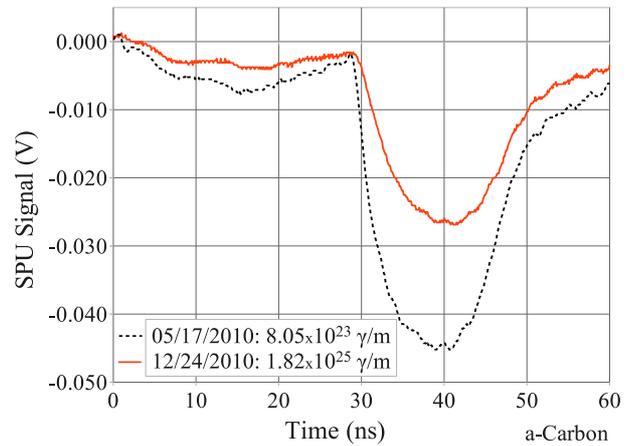}
   \vskip -1mm
   \caption{Comparison of two-bunch signals in May and December of 2010 in an
            amorphous-carbon-coated aluminum vacuum chamber shows a substantial
            reduction in cloud buildup. SBE signals from positron bunches of 
            population \mbox{$4.2{\times}10^{10}$} spaced by 28~ns
            were used for this purpose of comparison. The synchrotron radiation
            photon dose increased from
            \mbox{$8.05{\times}10^{23} \gamma$/m} to \mbox{$1.82{\times}10^{25} \gamma$/m}
            over these seven months.}
   \label{TwoBunchaC}
\end{figure}

In contrast, the cloud-producing properties of an
amorphous carbon coated chamber showed a strong dependence on radiation dose between May
and December of 2010, as shown in Fig.~\ref{TwoBunchaC}. The SBE signals were reduced
by about a factor of two for two 5.3~GeV bunches carrying \mbox{$4.2{\times}10^{10}$} 
positrons each, 28~ns apart. The integrated linear photon density increased from 
\mbox{$8.05{\times}10^{23} \gamma$/m} 
to \mbox{$1.82{\times}10^{25} \gamma$/m} over this period, since the chamber
had not been previously subjected to high-current running. 
The time dependence of the signals provides additional information 
on the nature of the conditioning effect. The signal from the second
bunch is much more sensitive to the secondary emission properties of the surface.
Since the signal of the leading bunch was reduced in similar proportion, 
seeding a much less dense cloud, we can deduce that the secondary yield properties 
did not change appreciably.
Indeed, full numerical simulations were consistent with a factor of two change
in the photoelectron production rate and with no change in secondary
yield~\cite{IPAC11:WEPC135,ECLOUD12:Fri1240}.

\section{Summary}
Time-resolved measurements of electron fluxes incident on the vacuum chamber wall in electron
and positron storage rings have been shown to be provide sensitivity to each of the various 
physical processes contributing to electron cloud buildup and decay. 
We have employed a simple technique of placing
an in-vacuum BPM-style button electrode behind a pattern of holes in the beam-pipe and 
digitizing the current signals obtained during and following the passage of a train of
beam bunches. The method provides information on the scattering of synchrotron radiation within the
pipe, the photoelectron production kinetic energy distribution, and the individual 
contributions of the various physical process contributing to secondary electron emission.
Accurate determinations of cloud lifetime have been obtained, as have
quantitative characterizations of photoelectron production and secondary emission properties of 
aluminum, amorphous carbon, diamond-like carbon and titanium-nitride coatings. The excellent
reproducibility of the measurements on a time scale of months has permitted the determination
of the beam-dose dependence of the surface properties of these electron cloud buildup mitigation
techniques. 

\section{Acknowledgments}
We wish to acknowledge contributions from the technical and administrative
staffs of the Cornell Laboratory for Accelerator-based ScienceS and Education.
This work is supported by the National Science Foundation 
under contract numbers PHY-0734867 and PHY-1002467 
and by the US Department of Energy under contract numbers
\mbox{DE-FC02-08ER41538} and \mbox{DE-SC0006505}.












\bibliographystyle{model1-num-names}
\bibliography{nim_sbe}
\end{document}